\documentclass{elsart5p}

\usepackage{graphics}
\usepackage{graphicx}
\usepackage{amssymb}
\begin{document}

\begin{frontmatter}


\title{\textbf{Magnetic properties and magnetocaloric effect of $\mbox{Tb}_2\mbox{Rh}_{3}\mbox{Ge}$}}

\author{Baidyanath Sahu} and 
\author{Andr\'{e} M. Strydom}

\address{Highly Correlated Matter Research Group, Department of Physics, University of Johannesburg, PO Box 524, Auckland Park 2006, South Africa}

\corauth[abd]{Corresponding author. baidyanathsahu@email.com}

\begin{abstract}

We report the structural, and magnetic properties as well as magnetocaloric effect of a polycrystalline compound of $\mathrm{Tb_2Rh_3Ge}$. This compound crystallizes with $\mathrm{Mg_2Ni_3Si}$-type of rhombohedral Laves phases (space group $R\overline{3}m$, hR18). The magnetic properties and magnetocaloric effect of $\mathrm{Tb_2Rh_3Ge}$ are explored through dc$\textendash$magnetization measurements. Temperature dependence of magnetization revealed that the compound exhibits ferromagnetic behavior with $T_C$$=$56 K. The field dependence of magnetization indicates that $\mathrm{Tb_2Rh_3Ge}$ is a soft ferromagnet. The obtained isothermal magnetic entropy changes ($\Delta S_m$) and refrigeration capacity (relative cooling power) for a change of magnetic field 0$\textendash$9 T are 12.74 J/kg$\textendash$K and 497(680) J/kg respectively. The Arrott plots and universal curve of normalized $\Delta S_m$ indicate that this compound undergoes a second order ferromagnetic phase transition.

\end{abstract}


\end{frontmatter}

\section{INTRODUCTION}

Ternary rare-earth based intermetallic compounds RTX (R = rare earth, T = transition metals and X = p block elements) have drawn considerable attention for their diversity of structural and magnetic properties. The interaction of localized magnetic moments arise from 4f$\textendash$electrons of R-ions and leads to a wealth of properties such as ferromagnetic, antiferromagnetic, and non$\textendash$magnetic behavior \cite{book1, book2, book3}. Recently, intermetallic compounds have been found attractive for the magnetocaloric effect (MCE) due to their large localized moments. The development of magnetic refrigeration technology based on the MCE is useful in place of gas for improving efficiency and for its environmentally friendly nature. Large MCE near room temperature would be favourable to utilise in industrial and household applications. However, the MCE in the cryogenic temperature domain is also desirable for some specific applications \cite{Book, review1, review2, fundamental1, fundamental2}. The investigation of magnetocaloric effect may have advantages for both the applications in magnetic refrigeration and the understanding of the fundamental properties of the material \cite{fundamental1, fundamental2, fundamental3, RCP1}. In recent years, investigation of new R$\textendash$based ternary intermetallic and oxide compounds has been found a resourceful approach to gain insight of the plausible design and composition to be used for profitable MCE materials, since they have potential applications at low temperature \cite{new,oxide}.

The MCE properties of rare earth based binary Laves phases in particular are of interest due to their diversity of magnetic properties and magnetocaloric effect \cite{Buschow,RCo2,RAl2,RNi2}. On the other hand, the literature revealed that a small number of ternary Laves phase compounds are known \cite{Y2Rh3Ge,R2Rh3Ga}. Rhomboheral Laves phase of rare-earth based ternary compounds are of interest for tremendous structural and magnetic properties \cite{Y2Rh3Ge,Ce2Rh3Ge,Pr2Rh3Ge,R2Rh3Ga,Gd2Rh3Ge}. $\rm{R_{2}Rh_{3}Ge}$ series of compounds with light rare earth elements such as Pr show first order ferro$\textendash$paramagnetic phase transition, whereas those with heavy rare earth elements, Gd and Er exhibit second order ferro$\textendash$paramagnetic phase transition \cite{Pr2Rh3Ge,Gd2Rh3Ge,Er2Rh3Ge}. In our previous studies, $\rm{Gd_{2}Rh_{3}Ge}$ and $\rm{Er_{2}Rh_{3}Ge}$ were found to show a considerable MCE effect \cite{Er2Rh3Ge}. In this paper, the structural, and magnetic properties, along with the magnetocaloric effect of a ternary rhomobohedral Laves phase compound of $\mathrm{Tb_2Rh_3Ge}$ are reported.

\section{EXPERIMENTAL DETAILS}
A polycrystalline sample of $\mathrm{Tb_2Rh_3Ge}$ was prepared by arc$\textendash$melting method. The stoichiometric amounts of the constituents (purities in 99.99 wt \%) were melted under high purity argon atmosphere on a water$\textendash$cooled copper$\textendash$hearth. The sample was flipped and remelted several times to ensure homogeneous mixing of the constituents. The obtained weight loss after final melting was less than 1~ wt.\%. The ingot was wrapped with tantalum foil and encapsulated in an evacuated quartz tube. The tube was kept for annealing at 1273 K for one week and then quenched in cool water. The phase purity of the sample was determined by collecting the room temperature powder X-ray diffraction (XRD) using Cu$\textendash$K$_\alpha$ radiation in a  Rigaku X-ray diffractometer. The diffraction pattern was analysed by the Rietveld refinement method using the ‘‘FULLPROF” software \cite{rietveld,fullprof}. The temperature and field dependence of the dc$\textendash$magnetization measurements were carried out using a Dynacool$\textendash$Physical Property Measurement System (PPMS) (made by Quantum Design, San Diego, USA) attached with a Vibrating Sample Magnetometer (VSM) option. Temperature dependence of the magnetization ($M(T)$) was carried out under zero field cooled (ZFC) and field cooled (FC) protocols \cite{Gd2Rh3Ge}.

\section{RESULTS AND DISCUSSIONS}
\subsection{Powder X$\textendash$ray diffraction}

\begin{figure}[!t]
	\includegraphics[scale=0.345]{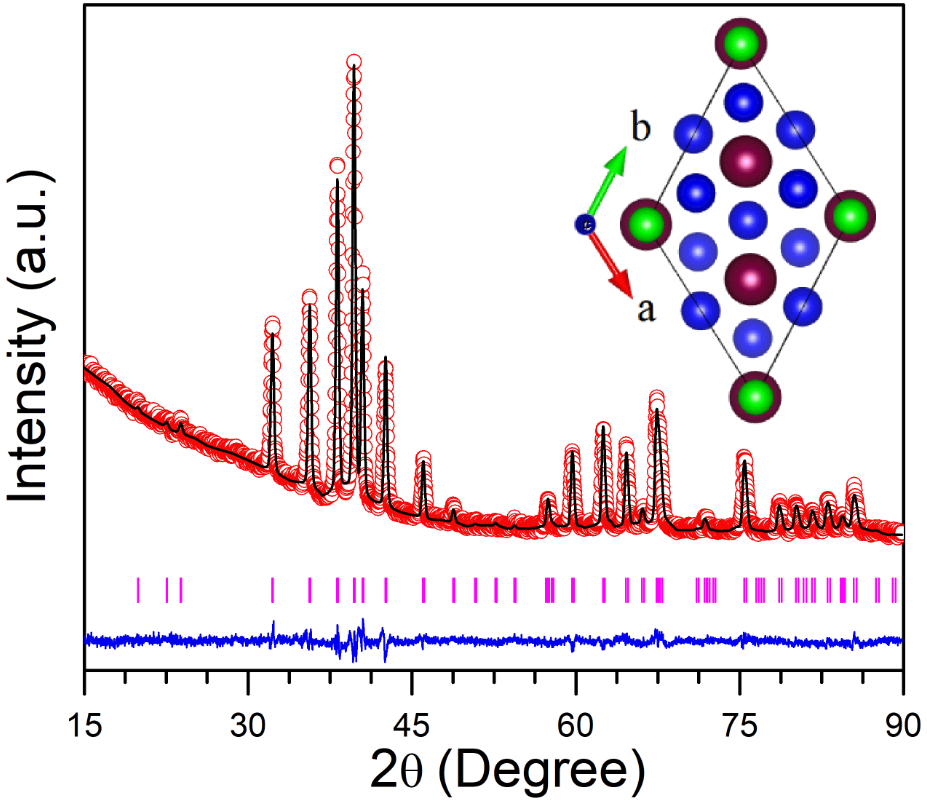}
	\caption{The powder XRD pattern of $\mathrm{Tb_2Rh_3Ge}$ along with Rietveld refinement fitting. The red circles indicate the experimentally observed intensity, and the black solid line is the calculated intensity assuming $R\overline{3}m$ space group. The difference curve is shown as a blue line and the location of the allowed Bragg peaks as vertical bars. Inset: A schematic of the rhombohedral crystal structure, which crystallizes in the $R\overline{3}m$ space group. The magenta spheres are Tb, blue are Rh and green are Ge.}
	\label{XRD}
\end{figure}

The recorded room temperature XRD patterns together with the Rietveld refinement fitting profile of $\mathrm{Tb_2Rh_3Ge}$ is shown in Fig.~\ref{XRD}. The Rietveld analysis was carried out using a rhombohedral phase with space group $R\overline{3}m$. The difference between the experimental and fitting data revealed that $\mathrm{Tb_2Rh_3Ge}$ crystallizes in a $\mathrm{Mg_2Ni_3Si}$$\textendash$type of rhombohedral Laves phases. The obtained refinement parameters are tabulated in Table~\ref{TABLE1}.

\begin{table}
\caption{\label{TABLE1} The lattice parameters and unit cell volume of $\mathrm{Tb_2Rh_3Ge}$ obtained from the Rietveld refinement of XRD patterns for the rhombohedral phase along with the atomic coordinate positions. The refinement quality parameter is $\chi^2$ = 3.40.}
\begin{tabular}{ccccc}\\ \hline \hline
\hspace{-1.5 in} a~=~b 	& \hspace{-0.0 in}  & \hspace{1.0 in} & \hspace{-0.1 in}  & \hspace{-0.5 in}5.556(2) \AA    \\
\hspace{-1.5 in} c  & \hspace{-0.0 in}  & \hspace{1.0 in} & \hspace{-0.1 in}  & \hspace{-0.5 in} 11.829(3) \AA \\
\hspace{-1.5 in} V  & \hspace{-0.0 in}  & \hspace{1.0 in} & \hspace{-0.1 in}  & \hspace{-0.5 in} 316.217(2)  \AA$^3$ \\
\hline
\hspace{0.1 in} Atomic coordinates  for $\mathrm{Tb_2Rh_3Ge}$ 	& \hspace{-0.2 in}  & \hspace{-0.2 in} & \hspace{-0.2 in}     & \hspace{-0.2 in} \\
\hline
\hspace{-1.5 in}Atom & \hspace{-2.0 in}Wyckoff & \hspace{-1.5 in}$x$ & \hspace{-1.2 in}$y$ & \hspace{-0.2 in}$z$ \\
 \hline
 & & & & \\
 \hspace{-1.5 in}Tb & \hspace{-2.0 in}$6c$ & \hspace{-1.5 in}0 & \hspace{-1.2 in}0 & \hspace{-0.2 in}0.3721(2)\\
 \hspace{-1.5 in}Rh & \hspace{-2.0 in}$9d$ & \hspace{-1.5 in}1/2 & \hspace{-1.2 in}0 & \hspace{-0.2 in}1/2\\
 \hspace{-1.5 in}Ge & \hspace{-2.0 in}$3a$ & \hspace{-1.5 in}0 & \hspace{-1.2 in}0 & \hspace{-0.2 in}0\\
 \hline \hline
\end{tabular}
\end{table}

\subsection{Magnetic  properties}

\begin{figure}[!t]
	\includegraphics[scale=0.32]{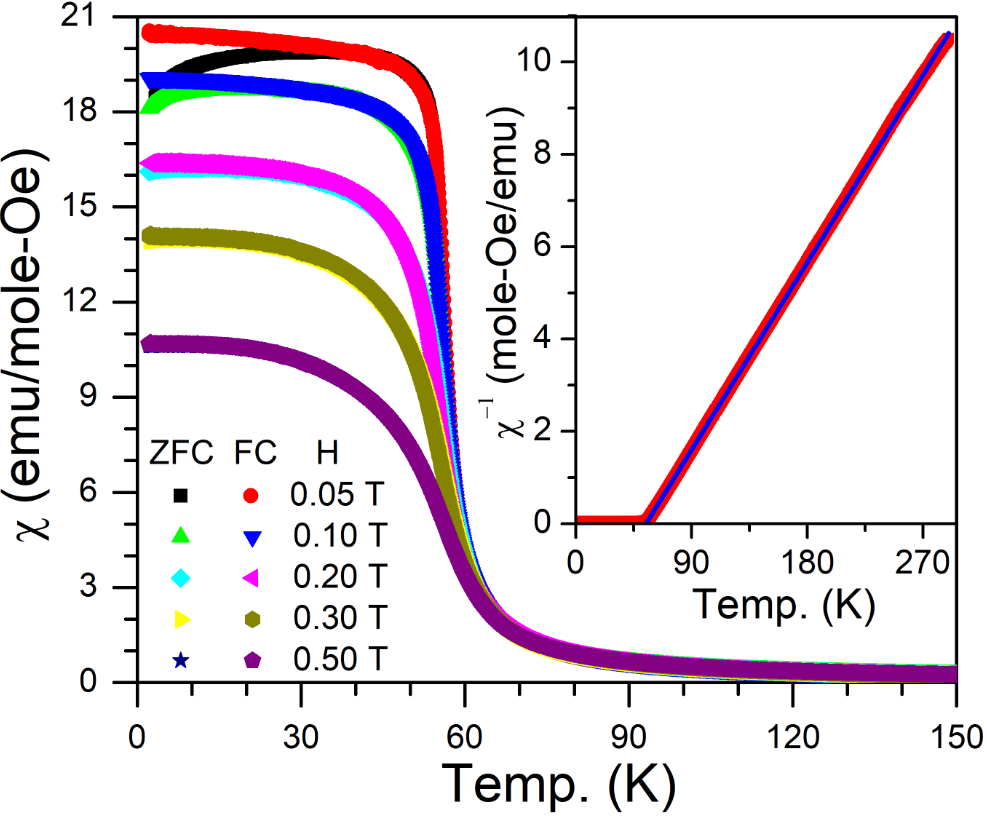}
	\caption{(a) Temperature dependence of ZFC and FC dc$\textendash$magnetic susceptibility 
			at different magnetic field for $\mathrm{Tb_2Rh_3Ge}$. Inset: Inverse dc$\textendash$magnetic susceptibility under 0.05 T, field cooling along with the Curie$\textendash$Weiss fitting.}
	\label{MT}
\end{figure}

Fig.~\ref{MT} displays the temperature dependence of ZFC and FC magnetic susceptibility ($\chi(T)$ = $M(T)/H$) measured under different applied magnetic fields. The ZFC and FC dc$\textendash$magnetic susceptibility curves for all fields show a typical paramagnetic to ferromagnetic transition. The Curie temperature ($T_C$) was derived from the peak value of d$\chi(T)$/d$T$~$vs.$~$T$ of FC curves (not shown in figure) and was found to be 56 K. Below $T_C$, the ZFC and FC curves show thermomagnetic irreversibility behaviour and is dependent on the applied magnetic field. Here, the decrease of ZFC magnetization with decreasing of temperature is the main cause of thermomagnetic irreversibility. This irreversibility behavior may result from the magnetocrystalline anisotropy or domain wall pinning effect \cite{Pr2Rh3Ge,book4}.

The reciprocal susceptibility $\chi^{-1}(T)$ of FC curve for the applied field of 0.05 T of $\mathrm{Tb_2Rh_3Ge}$ is shown in the inset of Fig.~\ref{MT}. The linear behavior of $\chi^{-1}(T)$ above 60 K follows the Curie$\textendash$Weiss law  $\chi^{-1}(T)$ = (${T-\theta_P}$)/C, where $\mathrm{C = N \mu^{2}_{eff}/3k_{B}}$ is known as the Curie constant and $\theta_P$ is the Weiss paramagnetic temperature. The Curie$\textendash$Weiss fit to the experimental data for $\mathrm{Tb_2Rh_3Ge}$ is shown as a blue line. The least$\textendash$squares (LSQ) fit of experimental data yielded the parameter $\theta_P$ = 54.5 K. This positive and large value of $\theta_P$ indicates the predominance of ferromagnetic exchange interaction in the compound. The effective moment ($\mathrm{\mu_{eff}}$) was estimated from the fitted value of Curie constant and is found $\mathrm{\mu_{eff}}$ = 9.42 $\mathrm{\mu_{B}}$/$\mathrm{Tb}$.  This obtained $\mathrm{\mu_{eff}}$ is slightly deviated from the theoretical value of a free $\mathrm{Tb^{3+}}$ ion, $\mathrm{g_{J}[J(J + 1)]^{1/2}}$ = 9.72 $\mathrm{\mu_{B}}$ for J = 6, and is attributed to the influence of the crystal electric field (CEF) effect. This result indicates that 4f $\textendash$shell electrons of $\mathrm{Tb^{3+}}$ ions are the predominant magnetic species in $\mathrm{Tb_2Rh_3Ge}$.

\begin{figure}[!t]
	\includegraphics[scale=0.32]{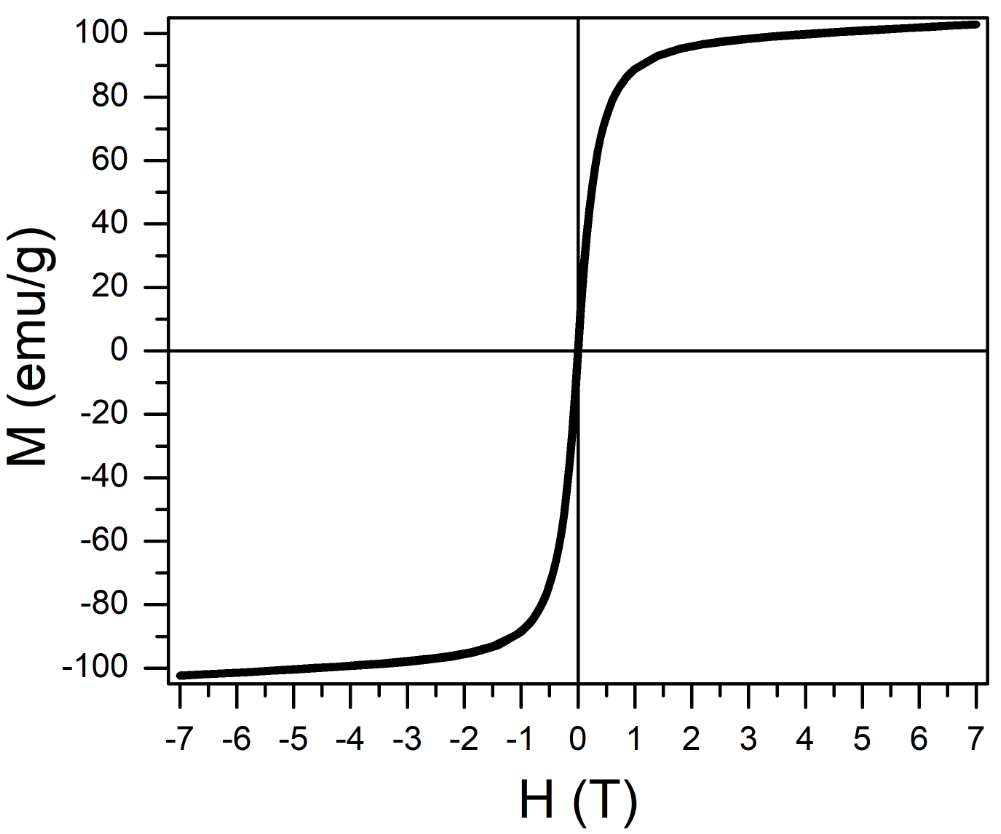}
	\caption{Magnetic field dependent magnetization loop at 2 K.}
	\label{2K}
\end{figure}

\begin{figure}[!t]
	\includegraphics[scale=0.32]{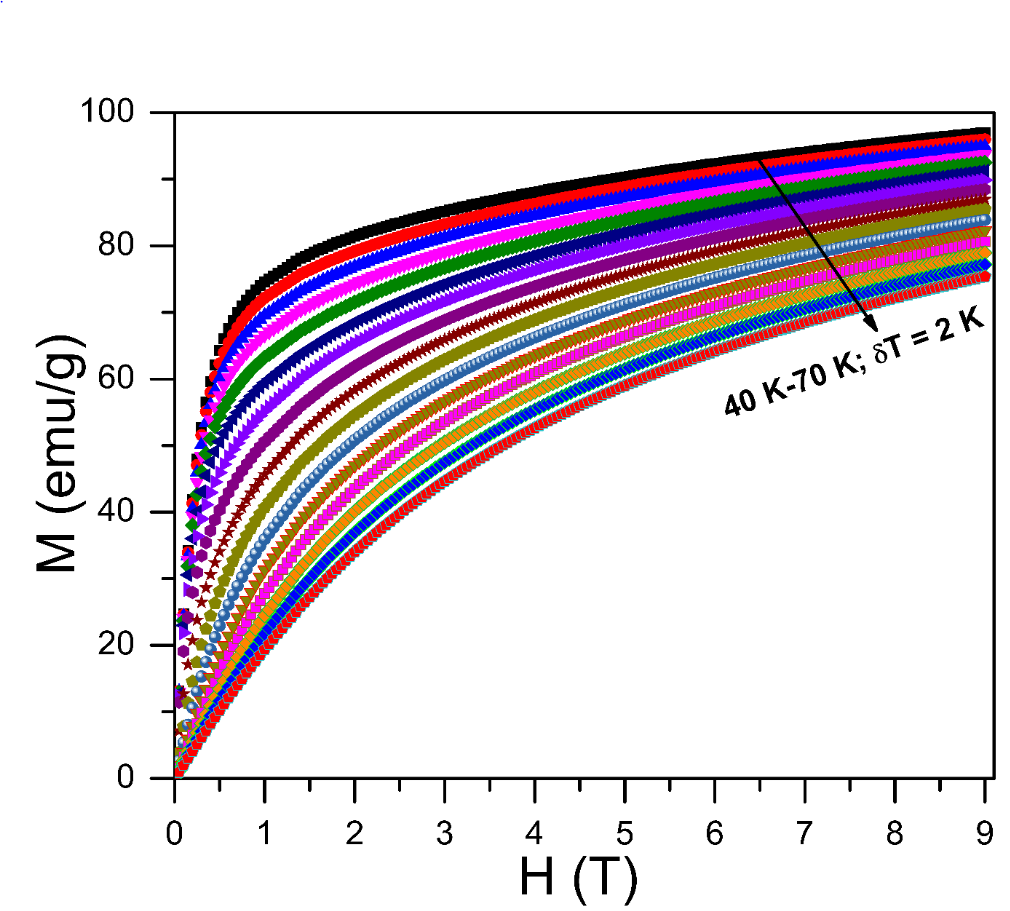}
	\caption{Isothermal magnetization at different temperatures ranging from 40 K to 70 K with a step of 2 K.}
	\label{MH}
\end{figure}

Fig.~\ref{2K} shows the field dependence magnetization loop at T = 2 K. The absence of detectable hysteresis loop at 2 K indicates that $\mathrm{Tb_2Rh_3Ge}$ is a soft ferromagnetic material. Soft ferromagnetic material is, in principle, a favourable attribute for applications in magnetic refrigerators \cite{Dy2Cu2In}. One can see that the high field magnetization part of Fig.\ref{2K} does not reach to full saturation even at 2 K. The obtained magnetization value at 2 K is 6.57 $\mu_B/\mathrm{Tb}$ (51.45(emu/g)/Tb) at 7 T, which is smaller than the spontaneous magnetic moment of a free $\mathrm{Tb^{3+}}$ ion (g$J$ = 9 $\mu_B/\mathrm{Tb}$). The reduction of the measured  saturation value may result from the magnetic anisotropy due to CEF effects \cite{Pr2Rh3Ge}. The magnetic isothermal ($M(H)$) of $\mathrm{Tb_2Rh_3Ge}$ was measured under maximum magnetic applied field up to 9 T, in a wide temperature range near the $T_C$. Fig.\ref{MH} shows the $M(H)$ at different temperatures from 40 to 70 K.  To avoid field cycling effect between two successive isothermal magnetizations, the sample was heated to 150 K after finishing the measurement of one isotherm. No metamagnetic signature is observed in the measured field range for $M(H)$ of Fig.~\ref{MH}. As seen from Fig.~\ref{MH}, the magnetization value at 9 T decreases from 97 emu/g to 75.4 emu/g by increasing temperature from 40 K to 70 K.

\begin{figure}[!t]
	\includegraphics[scale=0.315]{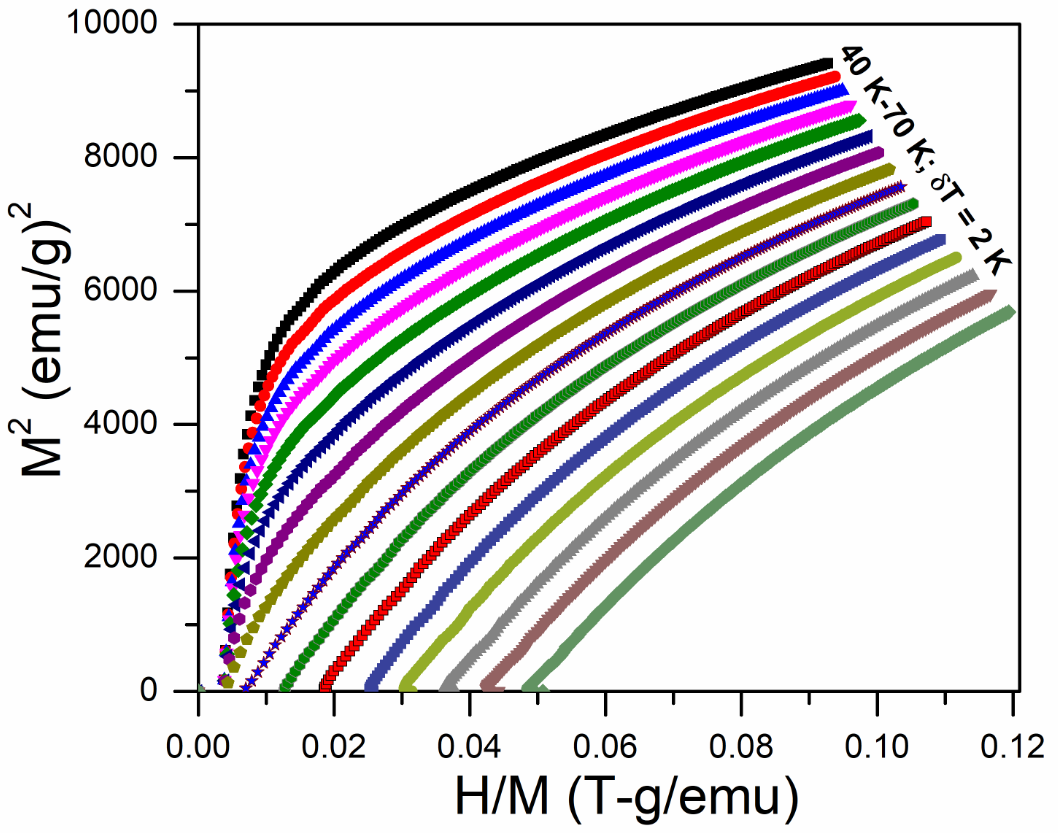}
	\caption{Arrott plots, $M^2~vs.~H/M$ at different temperatures from 40 K to 70 K with a step of 2 K.}
	\label{AP}
\end{figure}

The order of magnetic phase transition was determined from $M(H)$ by performing Arrott plots ($M^2~vs.~H/M$). Fig.~\ref{AP} shows the $M^2~vs.~H/M$ in the temperature range of 40 K to 70 K. As per the Banerjee criterion \cite{Banarjee}, the positive slope in $M^2~vs.~H/M$ indicates that $\mathrm{Tb_2Rh_3Ge}$ undergoes second order ferromagnetic phase transition, whereas a negative slope is expected at a first order magnetic transition. The order of magnetic phase transition is also confirmed by employing universal scaling plots of normalized magnetic entropy change and is discussed in the next section.

\subsection{Magnetocaloric effect}
\begin{figure}[!t]
	\includegraphics[scale=0.325]{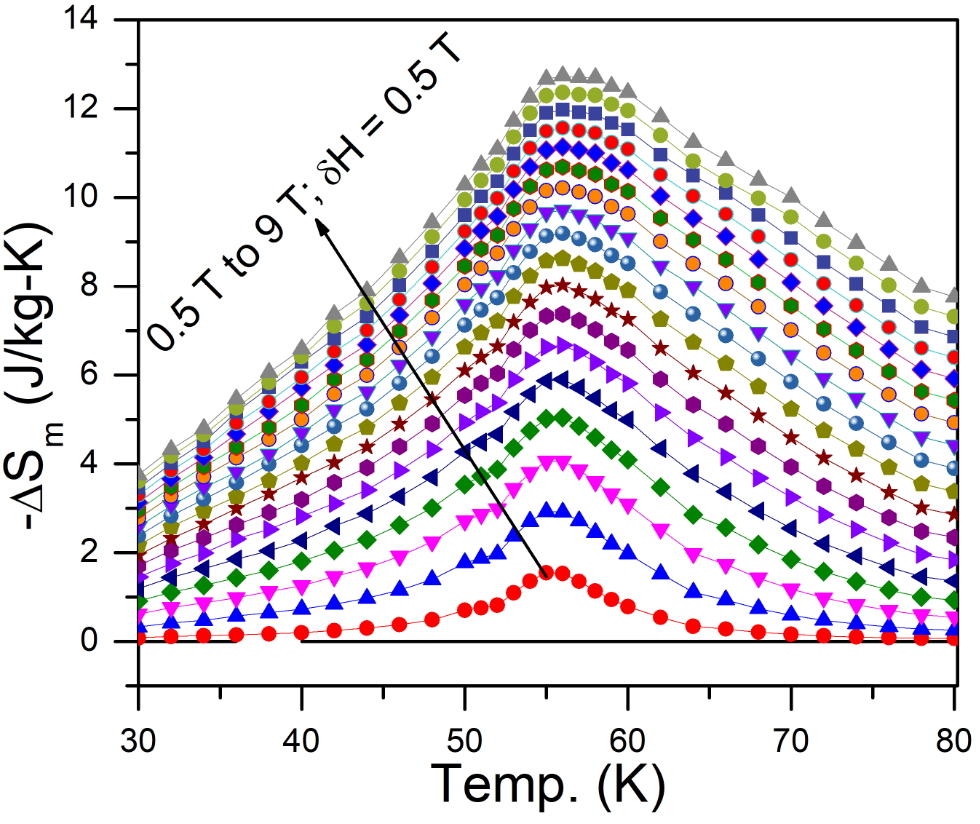}
	\caption{The isothermal magnetic entropy change as a function of temperature for various
	magnetic field changes $\Delta H$ up to 0$\textendash$9 T for $\mathrm{Tb_2Rh_3Ge}$.}
	\label{MCE}
\end{figure}

The isothermal magnetic entropy change ($\Delta S_m$) of $\mathrm{Tb_2Rh_3Ge}$ is calculated from the isothermal magnetization curves by using the following Maxwell's magnetic thermodynamic relation \cite{Book};
\begin{equation}
\Delta S_m (T, H) = \int_{H_i}^{H_f} \left(\frac{\partial M}{\partial T}\right)_H dH'
\label{DeltaS}
\end{equation}
where $H_i$ is the initial magnetic field (zero in these experiments)
and $H_f$ is the final magnetic field. The calculated temperature variation of $- \Delta S_m$ for several values of applied fields in the range of 0 T to 9 T is shown in Fig.\ref{MCE}. It is found that the $- \Delta S_m$ value is positive for a wide range of temperature with a single broad peak, indicating ferromagnetic$\textendash$paramagnetic transition. The maximum values of $- \Delta S_m$ appears around the $T_C$ and is found to increase with increasing magnetic field up to 9 T. This indicates that large magnetic field can produce large values of $- \Delta S^{max}_m$ \cite{RE2CuSi3}. The obtained maximum values of $- \Delta S^{max}_m$ are 5.04, 9.19, 11.14 and 12.74 $\mathrm{J/kg-K}$ for the change of magnetic field 0$\textendash$2, 0$\textendash$5, 0$\textendash$7 and 0$\textendash$9 T. 

The refrigeration capacity (RC) and/or relative cooling power (RCP) are also considered as alternative parameters to quantify the heat transfer between hot and cold reservoirs. The RC is determined by the formula; $RC~=~\int_{T_1}^{T_2} |\Delta S_m| dT$. Here, $T_1$ and $T_2$ are the temperatures below and above $T_C$, respectively, at the half maximum of the $\Delta S_m$ peak in an ideal thermodynamic cycle. RCP is defined as the product of the maximum magnetic entropy change $|- \Delta S^{max}_m|$ and the full width at half maximum ($\delta T_{FMHM})$ of  $\Delta S_m$ $vs.$~$T$ curves; RCP~$=$~$|- \Delta S^{max}_m|$~$\times$~$\delta T_{FMHM}$ \cite{Book,RCP1,RCP}. It is observed that the evaluated RC and RCP values gradually increase with the increasing of change in applied magnetic fields. The RC(RCP) values are 80(108), 240(320), 360(480) and 497(680) J/kg for a change of magnetic field 0$\textendash$2, 0$\textendash$5, 0$\textendash$7 and 0$\textendash$9 T respectively. For comparison, the magnetic transition temperature, $- \Delta S_m$, the RC and RCP values for the change of magnetic field 0$\textendash$5 T of $\mathrm{Tb_2Rh_3Ge}$ as well as the values for other related MCE magnetic compounds are listed in Table~\ref{tableMCE}. As seen from this table, the MCE results of $\mathrm{Tb_2Rh_3Ge}$ are comparable and even large from reported MCE compounds. This comparison suggests that $\mathrm{Tb_2Rh_3Ge}$ belongs to a class of considerable MCE materials.

\begin{figure}[!t]
	\includegraphics[scale=0.315]{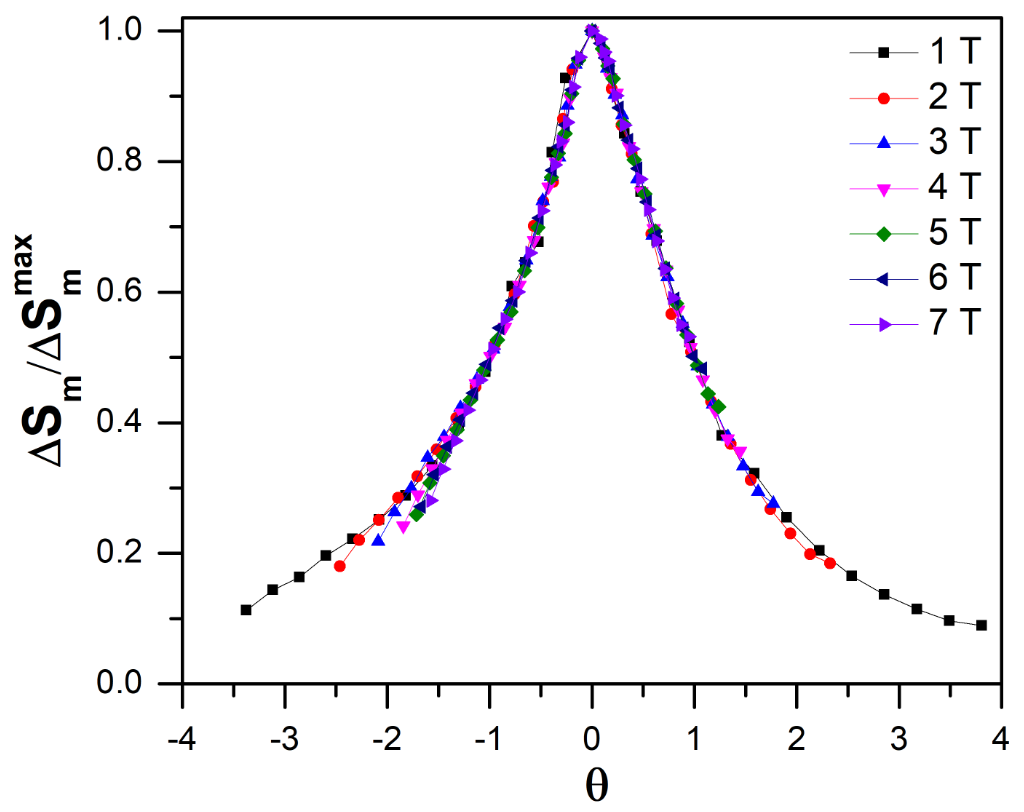}
	\caption{Universal scaling plot by normalized magnetic entropy change $\Delta S_m$/$\Delta S^{max}_m$ as a function of the	rescaled temperature $\theta$ around $T_C$ for different values of magnetic field of $\mathrm{Tb_2Rh_3Ge}$.}
	\label{scaling}
\end{figure}

\begin{table}[!t]
\caption{\label{tableMCE} The transition temperature ($T_M$), the magnetic entropy change ($\Delta S_m$) and relative cooling power (RCP)/refrigeration capacity (RC), under a magnetic field change of 0$\textendash$5~T for $\mathrm{Tb_2Rh_3Ge}$ together with other related MCE compounds.}
\begin{tabular}{cccccc} \\ \hline
\hspace{-0.05 in}Compound	 & 	\hspace{-0.1 in} $T_M$	 & 	\hspace{0.12 in}$-\Delta S_m$	 &	 \hspace{0.12 in}RCP	 & \hspace{0.12 in}RC	 &		 \hspace{0.12 in}Ref  \\
\hspace{-0.05 in} & \hspace{-0.1 in} (K) & \hspace{0.12 in} (J/kg-K) & \hspace{0.12 in} (J kg$^{-1}$) & \hspace{0.12 in} (J kg$^{-1}$) &  \hspace{0.12 in}   \\ \hline	
\hspace{-0.05 in}$\mathrm{Gd_2Rh_3Ge}$ & \hspace{-0.1 in} 64 & \hspace{0.12 in} 9.4 & \hspace{0.12 in} 352 & \hspace{0.12 in} --- & \hspace{0.12 in} \cite{Er2Rh3Ge}  \\
\hspace{-0.05 in}$\mathrm{Er_2Rh_3Ge}$ & \hspace{-0.1 in} 21 & \hspace{0.12 in} 9.2 & \hspace{0.12 in} 225 & \hspace{0.12 in} --- &  \hspace{0.12 in} \cite{Er2Rh3Ge}  \\
\hspace{-0.05 in}$\mathrm{Tb_2CoAl_3}$ & \hspace{-0.1 in} 47 & \hspace{0.12 in} 8.6 & \hspace{0.12 in} 368 & \hspace{0.12 in} --- & \hspace{0.12 in} \cite{Tb2CoAl3}  \\
\hspace{-0.05 in}$\mathrm{TbPdAl}$ & \hspace{-0.1 in} 43 & \hspace{0.12 in} 11.4 & \hspace{0.12 in}  --- & \hspace{0.12 in} 350 &  \hspace{0.12 in} \cite{TbPdAl}  \\
\hspace{-0.05 in}$\mathrm{TbPtMg}$ & \hspace{-0.1 in} 58 & \hspace{0.12 in} 5.1 & \hspace{0.12 in} 192 & \hspace{0.12 in} ~142 & \hspace{0.12 in} \cite{TbPtMg}  \\
\hspace{-0.05 in}$\mathrm{TbCo_2}$ & \hspace{-0.1 in} 231 & \hspace{0.12 in} 6.9 & \hspace{0.12 in} 357 & \hspace{0.12 in} --- & \hspace{0.12 in} \cite{TbCo2}  \\
\hspace{-0.05 in}$\mathrm{TbNiIn}$ & \hspace{-0.1 in} 71 & \hspace{0.12 in} 5.3 & \hspace{0.12 in} --- & \hspace{0.12 in} 191 & \hspace{0.12 in} \cite{TbNiIn}  \\
\hspace{-0.05 in}$\mathrm{Tb_2Rh_3Ge}$ & \hspace{-0.1 in} 56 & \hspace{0.12 in} 9.2 & \hspace{0.12 in} 320 & \hspace{0.12 in} 240 & \hspace{0.12 in} PW  \\
\hline
PW: Present Work	
\end{tabular}	
\end{table}

Here, the universal curve of magnetic change entropy has been employed to confirm the order of the
magnetic phase transition~\cite{bonilla2010new,bonilla2010universal,franco2008universal,franco2010scaling,franco2006scaling}. Universal curves are constructed by the normalized entropy change ($\Delta S_m$/$\Delta S^{max}_m$) against rescaled temperature ($\theta$). $\theta$ is evaluated by the expression;  
 \begin{eqnarray} \nonumber
 \theta = (-(T - T_C)/(T_{r1} - T_C)), T<T_C \\
 ((T - T_C)/(T_{r2} - T_C)), T>T_C,
 \end{eqnarray}
where $T_{r1}$ and $T_{r2}$ are the temperatures corresponding to 50$\%$ of $\Delta S^{max}_m$ below and above $T_C$, respectively for each applied field. Fig.~\ref{scaling} shows the universal scaling plots ($\Delta S_m$/$\Delta S^{max}_m$~$vs.$ $\theta$) of $\mathrm{Tb_2Rh_3Ge}$. It is to be noted that the $\Delta S_m$/$\Delta S^{max}_m$~$vs.$ $\theta$ curves nearly collapse onto one master curve for all fields, indicating a second order magnetic phase transition in $\mathrm{Tb_2Rh_3Ge}$.

\section{CONCLUSIONS}

We have successfully synthesized a polycrystalline compound $\mathrm{Tb_2Rh_3Ge}$, and determined that it crystallizes in $\mathrm{Mg_2Ni_3Si}$$\textendash$type of rhombohedral Laves phase. The magnetic properties, including order of the magnetic phase transition, and magnetocaloric effect have been systematically investigated. Dc$\textendash$magnetization results revealed that $\mathrm{Tb_2Rh_3Ge}$ is a soft ferromagnet with the magnetic transition temperature, $T_C$ = 56 K. The second order magnetic phase transition in $\mathrm{Tb_2Rh_3Ge}$ is confirmed from both the Arrott plots and the universal curve of rescaling entropy change under different magnetic fields. A considerable reversible MCE is observed in $\mathrm{Tb_2Rh_3Ge}$. The obtained value of $- \Delta S^{max}_m$ is 12.74 J/kg$\textendash$K for the magnetic field change of 0$\textendash$9 T and the corresponding RC (RCP) is 497 (680). The present results may provide interest for investigating the MCE among members of the searching new $\mathrm{R_2T_3X}$ series of compounds.

\section*{ACKNOWLEDGEMENT}   
This work is supported by Global Excellence and Stature (UJ-GES) fellowship, University of Johannesburg, South Africa. AMS thanks the URC/FRC of UJ and the SA NRF (93549) for financial assistance.

\end{document}